# The Trapping and Characterization of a Single Hydrogen Molecule in a Continuously Tunable Nanocavity

Hui Wang[1,2,+], Shaowei Li[2], Haiyan He[2,3,+], Arthur Yu[2], Freddy Toledo[4], Zhumin Han[2], W. Ho[2,4,*] and Ruqian Wu[1,2,*]

[1]*State Key Laboratory of Surface Physics and Department of Physics, Fudan University, Shanghai 200433, CHINA*

[2]*Department of Physics and Astronomy, University of California, Irvine, CA 92697-4575, USA*

[3]*Department of Physics, University Science and Technology of China, Hefei, Anhui 230026, CHINA*

[4]*Department of Chemistry, University of California, Irvine, CA 92697-2025*

ABSTRACT: Using inelastic electron tunneling spectroscopy with the scanning tunneling microscope (STM-IETS) and density functional theory calculations (DFT), we investigated properties of a single $H_2$ molecule trapped in nanocavities with controlled shape and separation between the STM tip and the Au (110) surface. The STM tip not only serves for the purpose of characterization, but also is directly involved in modification of chemical environment of molecule. The bond length of $H_2$ expands in the atop cavity, with a tendency of dissociation when the gap closes, whereas it remains unchanged in the trough cavity. The availability of two substantially different cavities in the same setup allows understanding of $H_2$ adsorption on noble metal surfaces and sets a path for manipulating a single chemical bond by design.

---

[+]These two authors contributed equally to this work.

Control of chemical reaction one molecule at a time has been a long sought after goal for chemists.[1, 2] For the ensemble system, chemical reactions have been well controlled by tuning the reaction conditions, such as temperature, pressure, catalyst, or electric potential. Even in nano-reactors, it is still a major challenge to cleave chemical bonds by design and the performance of nano-catalysts strongly depends on their dimension and geometry.[3] The Scanning Tunneling Microscope (STM) is an ideal tool for the manipulation and characterization of chemical bonds in the nano-cavity that is formed by the STM tip and substrate and contains a single molecule.[4-8] The geometry and size of STM nanocavities can be precisely tuned by changing the substrate topography and the tip-substrate distance. Tunneling electrons can provide the driving force to induce atomic and molecular motions leading to the dissociation and formation of chemical bonds.[9] The electronic, vibrational, and structural properties of the trapped molecule can be probed by spectroscopic STM.[10]

Hydrogen molecule is involved in numerous chemical reactions. It is commonly applied to reduce organic compounds in multiple fields ranging from pharmaceutical, electronics, to green energy applications.[11-13] As the smallest molecule, $H_2$ is ideal for probing the properties of nano-reactors. Previous STM studies have revealed that the adsorption of $H_2$ on metal surfaces is extremely weak, mainly via the van der Waals forces.[14] Rotational and vibrational modes of $H_2$ have been detected on Au (110) and other surfaces[15, 16] using inelastic electron tunneling spectroscopy (IETS). It was found that the bond length of $H_2$ slightly expands as the gap of the STM nano-cavity was decreased.[15] Obviously, further studies of the changes in the rotational and vibrational modes with the tip-substrate distance (i.e., the size of cavity) offer unique opportunities to perceive with precise control the alternation of intra- and inter-molecular bonds in different chemical environments.

Au (110) has 2×1 missing row reconstruction and hence may form two different nano-cavities when the tip is parked above a Au row or over the trough between Au rows, as illustrated in Fig. 1(a). Therefore, we may explore the effect of the size and shape of nanocavities on chemical bonds. The hydrogen molecules are freely translating and rotating on the surface when dosed at 10 K. The experiments were performed using a home-built STM operating at 10 K and $3\times10^{-11}$ Torr.[17] The Au(110) 2x1 reconstructed surface was prepared by cycles of $Ne^+$ sputtering and annealing up to 720 K. The electrochemically etched silver tip is used. The $d^2I/dV^2$ spectra were recorded by monitoring the second harmonic output of the lock-in amplifier. The bias was applied to the sample with the tip connected to the current amplifier at virtual ground, and was swept and modulated at 5 mV and 345 Hz. The tip-substrate separation was determined by keeping the set point current constant at 2 nA while changing the set point bias with feedback on.

The topographic images taken after dosing molecular hydrogen reveal single atom resolution over the Au rows [Fig 1(d-e)], while the resolution over the trough is almost unchanged. The STM-IETS measurements over the Au rows (denoted by ta-cavity below) show two groups of inelastic excitation signals. As shown in Fig. 1(b), the strong signal around 13 mV is attributed to the vibrational motion of $H_2$ bouncing between the tip and Au rows according to the results of density functional theory (DFT) calculations, while the weaker signal around 42 mV is the j=0 to j=2 rotational excitation of $H_2$.[15] When the tip approaches the Au rows, the vibrational mode shifts to higher energy, whereas the rotational mode shifts to lower energy. An additional mode around 9 mV appears at the closest tip-substrate distance [bottom spectrum of Fig. 1(b)], which is assigned to the hindered translational mode of the trapped $H_2$. Using the 3D rigid free rotor model, we found that the energy shift of the rotational mode from 43 to 41 meV corresponds to an expansion of the H-H bond length from 0.746 Å to 0.766 Å. The experimental data indicate that the H-Au interaction is enhanced and the H-H bond weakens as the dimension of the tip-atop nano-cavity decreases. In contrast, the measurements for the tip positioned over the trough (denoted by tt-cavity below) reveal very different features in the spectra. The IETS spectra reveal the bouncing vibrational mode of $H_2$ at 5 meV and a rotational mode at 42 meV. These two modes exhibit no resolvable shift when the tip-substrate distance shrinks, while the weaker hindered translational mode around 18 meV shows non-monotonic shift as depicted in Fig. 1(c).

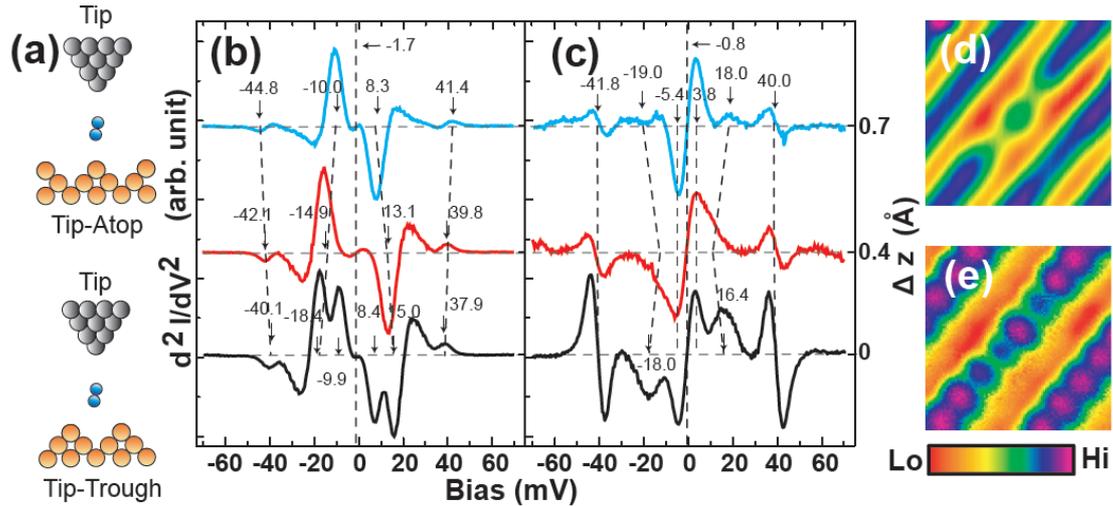

**Figure 1.** (a) Schematic diagram of two types of nano-cavities when tip is above a Au row (Tip-Atop) or over a trough (Tip-Trough). (b) STM-IETS measurements of a $H_2$ molecule trapped in the Tip-Top nano-cavity for three tip-substrate separations. The bouncing vibrational mode energy shifts up while the rotational excitation energy shifts down as the tip moves closer to the surface. (c) STM-IETS measurements of a $H_2$ molecule trapped in the Tip-Trough nano-cavity. The vibrational and rotational excitations do not reveal shift in energy for different tip-substrate distances. The center of the IETS spectra in (b) is around -1.7 mV while -0.8 mV in (c). The difference in the spectral offsets is possibly due to the work function difference between the atop and trough rows. The vibrational and rotational energies are determined from the average of the IETS features at the positive and negative biases of the spectra. (d-e) Topographic images were taken over the same area without (d) and with (e) a trapped hydrogen molecule. The one-atom and two-atom defects in a Au row can be clearly resolved when scanning with a trapped $H_2$.

Physical insights for understanding the enhanced spatial resolution of the Au rows and variations in energy of the hydrogen vibrational and rotational modes were obtained from DFT calculations using the Vienna Ab-initio Simulation Package (VASP).[18-20] We used the projector augmented plane wave (PAW) method for the description of core-valence interactions.[21-24] Spin-polarized generalized gradient approximation (GGA) with the Perdew-Burke-Ernzerhof (PBE) functional was employed to represent the exchange and correlation interactions among valence electrons (H-1s and Au-5d6s).[25] For the correct description of the physisorption of $H_2$ on Au (110) 2×1, the van der Waals correction was invoked in the self-consistent loop, using the non-local optB86b-vdW functional that was implemented in VASP.[26,

[27] The energy cutoff of the plane-wave expansion was as large as 700 eV in order to achieve a high quality convergence according to our test calculations.

As shown in Fig. 2, we explored eight high symmetry adsorption geometries of $H_2$ in the ta- and tt-cavities. We used a slab with 5 Au layers to simulate the Au surface and a Au pyramid to represent the STM tip. To mimic the environment for the adsorption of a single $H_2$ molecule, we adopted a (2x3) supercell in the lateral plane and fixed its size according to the optimized lattice constants for the bulk cubic Au (a = b = c = 4.17 Å). The two bottommost Au layers in the substrate and the topmost layer of Au atoms in the tip were frozen at their bulk-like positions, while other Au atoms and the $H_2$ molecule were fully relaxed according to the calculated atomic forces. The 11×11×1 Monkhorst-Pack[28] k-meshes were used to sample the Brillouin zone. The criterion for structural optimization was that the atomic force on each atom became less than 0.01 eV/Å and the energy convergence was better than $10^{-7}$ eV.

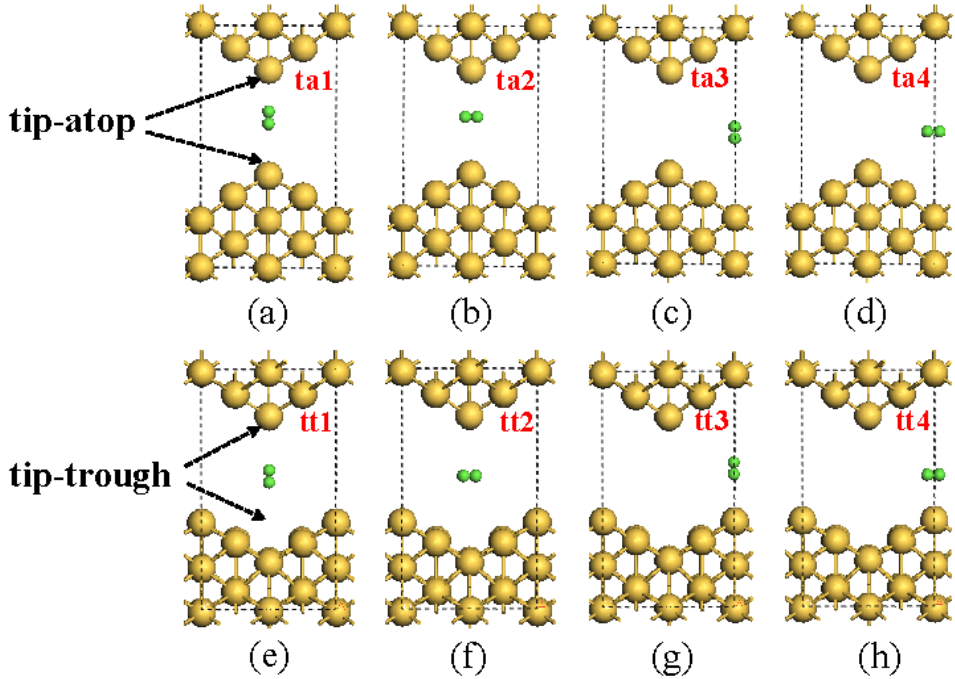

**Figure 2.** Schematic diagram of the adsorbed $H_2$ molecule in different STM nano-cavities: (a)-(d) Different adsorption sites of $H_2$ molecule in ta-cavity; (e)-(h) Different adsorption sites of $H_2$ molecule in tt-cavity. Yellow and green balls represent Au and H atoms, respectively.

We use the binding energy per $H_2$ molecule, defined as $E_b = E_{H_2/Cavity} - E_{Cavity} - E_{H_2}$, to describe the strength of $H_2$-Au interaction. Here $E_{H_2/Cavity}$ is the total energy of the cavity with a H$_2$ molecule adsorbed in it; $E_{Cavity}$ and $E_{H_2}$ denote the total energies of the cavity and the free H$_2$ molecule, respectively. From the binding energy curves in Fig. 3(a), it can be seen that the H$_2$ molecule prefers the trough site on the Au (110) surface when the tip is far away, with a binding energy of −89 meV in a vertical geometry (cf. ta3 and tt1 in Fig. 2). The preferential adsorption site of the H$_2$ molecule shifts to the Au row when the tip-substrate distance becomes smaller than 8 Å. The H-H bond length also expands, which leads to the reduction of rotational energy, as illustrated in Fig. 3(b). The magnitude of E$_b$ for H$_2$ in the ta-cavity increases rapidly and reaches its maximum at d≈5 Å, as depicted in Fig. 3(a). We find that the horizontal adsorption geometry (e.g., ta2) eventually converts to the vertical adsorption geometry, ta1, during structural optimization when the tip-substrate distance is smaller than 5.7 Å. It appears that the presence of the STM tip over the row of Au atoms establishes an attractive environment for H$_2$ molecules on Au (110), since the magnitude of E$_b$ (>300 meV) is much larger than those on the clean Au (110) surface. Therefore, a H$_2$ molecule can be trapped in the tip-substrate junction for a long time for experimental studies at low temperatures, which explains the enhanced spatial resolution in Fig. 1(e). The H$_2$ molecule is displaced from the center of the cavity when tip-substrate distance is smaller than 5 Å. Strikingly, the bond length of H$_2$ in the ta1 configuration may expand to 0.88 Å at around d ~ 5.0 Å, indicating that the H$_2$ molecule is ready to dissociate in the small ta-cavity. In contrast, the binding energy of H$_2$ in the tt-cavity remains small in the entire range of gap distance. Furthermore, the bond length of H$_2$ barely changes as the tt-cavity shrinks, which is mainly due to the fact that H$_2$ has room to adjust its position within the trough of Au surface. These results show that the binding energy and the structure of the H$_2$ molecule are sensitive to the variation in the tip-substrate separation over the row of Au atoms. This sensitivity leads to the ability to resolve the variation of gap size when the tip is positioned over the Au rows. In contrast, the binding energy and structure of H$_2$ in the trough cavity are insensitive to the change of tip-substrate distance due to less lateral confinement for the hydrogen.

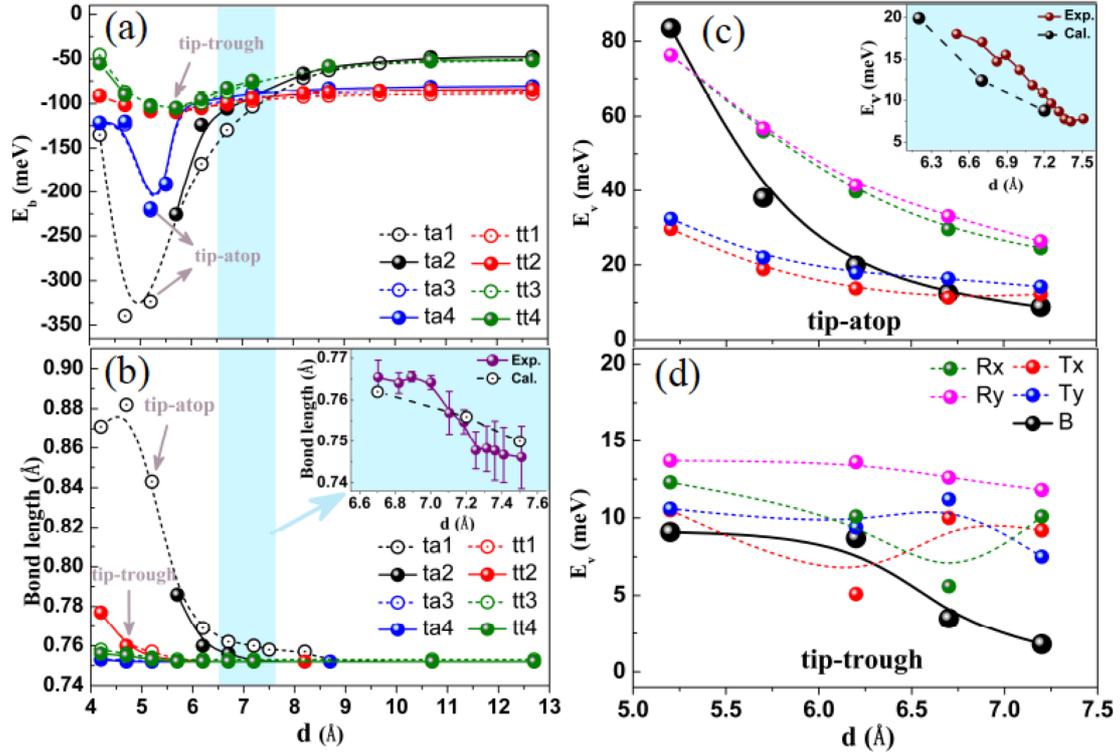

**Figure 3.** (a) Binding energy and (b) bond length versus variation of the tip-substrate distance for $H_2$ molecule at different adsorption sites sketched in Fig. 2 (a)-(h). The inset in (b) shows the experimentally derived bond length versus tip-substrate distance in the range 6.6 Å to 7.6 Å that can be directly compared to the calculated results in the light blue shaded area. Vibrational energies, $E_v$, versus tip-substrate distance for $H_2$ molecule adsorbed in (c) tip-atop junction and (d) tip-trough junction. The inset in (c) shows the experimental vibrational energies for tip-substrate distance in the range from 6.2 Å~ 7.6 Å that can be directly compared to the calculated results.

To compare with the experimental data in Fig. 1, we plotted the d-dependent energies of the five $H_2$ vibrations (the stretch mode has a much higher energy and is not shown here), namely, the bouncing mode (B), hindered rotation modes ($R_x$ and $R_y$), and hindered translation modes ($T_x$ and $T_y$), in Figs. 3(c) and 3(d). In the ta1 geometry, all the vibrational energies increase monotonically with the reduction of d. In particular, the energy of the bouncing mode increases from 8.8 to 19.9 meV as the tip-substrate gap shrinks from 7.2 to 6.2 Å, and a good agreement with the experimental data is shown in the inset of Fig. 3(c). The bouncing mode and the hindered translational modes split in energy as the tip-substrate gap decreases, which allows them to be experimentally resolved. Therefore, the observed vibrational mode around 13 mV in Fig. 1(b) can be assigned to the bouncing mode of the $H_2$ molecule trapped in the ta-cavity, while the new IETS feature at small tip-substrate separation can be assigned to the hindered translational modes. The rapid frequency increase of the bouncing mode for reduced d suggests an enhancement of the H-Au interaction at both ends of $H_2$. In contrast, the bouncing mode in the tt2 geometry is very soft for d > 6.5 Å as shown in Fig. 3(d), in line with the small binding energy for this geometry. Moreover, all the vibrational modes have energies in the range 5-13 meV and show a weak dependence on d. The hindered translational modes show non-monotonic shifts when d changes for the tip-trough geometry, which is also consistent with the experimental observations shown in Fig. 1(c).

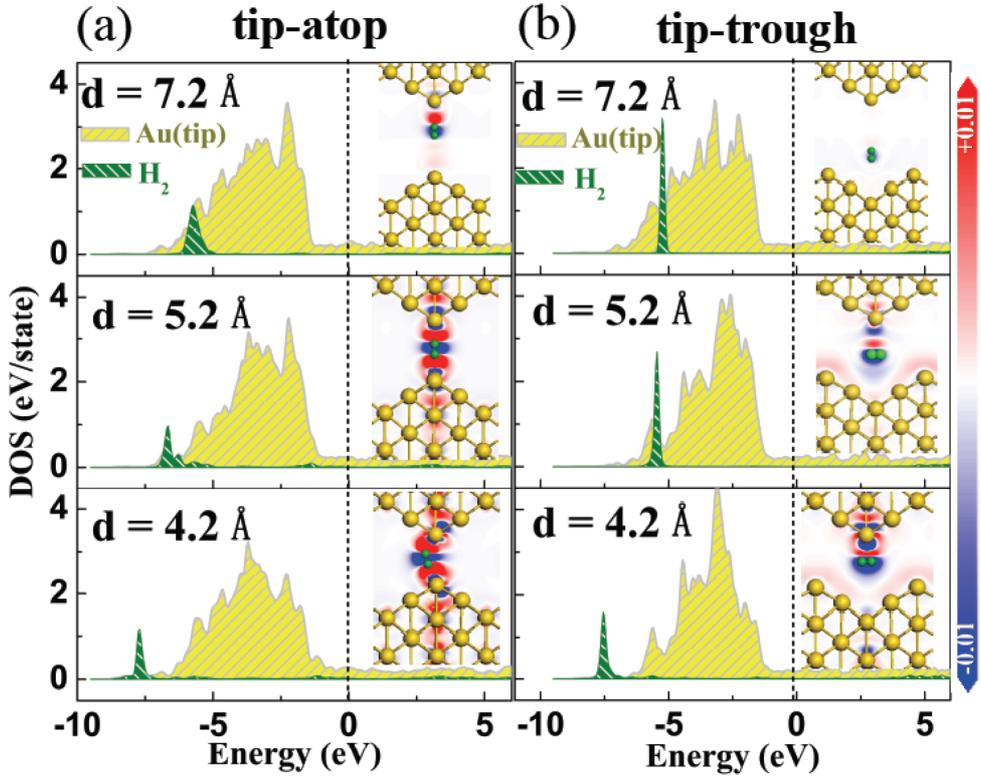

**Figure 4.** Local density of states (LDOS) of adsorbed $H_2$ molecule and Au (tip) atom in (a) tip-atop junction and (b) tip-trough junction with tip-substrate distance of 7.2 Å, 5.2 Å and 4.2 Å. The corresponding charge difference density is given in each panel: red and blue represent the charge accumulation and depletion at 0.01 e/Å$^3$, respectively.

Furthermore, we presented in Fig. 4 the local density of states (LDOS) of $H_2$ molecule at three tip-substrate distances: 7.2 Å, 5.2 Å, and 4.2 Å, along with the LDOS of the Au tip atom that directly interacts with $H_2$. Small but visible intermixes between H and Au states around the Fermi level can be found for the ta-geometry but not for the tt-geometry, corresponding to the slow change of vibrational properties as observed in Fig. 1(c) and Fig. 3(d). The main change is that the $H_2$-1s/Au-5d resonant peak is gradually pushed down from -6 eV to -8 eV, as the tip-substrate distance decreases from 7.2 Å to 4.2 Å. Note that the $H_2$-1s peak splits off from the Au-5d bands for d < 5.2 Å. The hybridization between $H_2$ and Au is hence reduced thereafter and a repulsive interaction between them develops in an ultra small tip-atop gap. From the plots of the charge redistribution in the insets of Fig. 4(a), the accumulation of charge between H and Au atoms (red region) increases as tip-substrate distance decreases from 7.2 Å to 5.2 Å, accompanied by charge depletion away from $H_2$ (blue region). When d shrinks further to 4.2 Å, the $H_2$ molecule is squeezed out of the junction and the charge rearrangement is somewhat reduced compared to that at d=5.2 Å. The LDOS curves of the $H_2$ molecule in the tt-cavity also show similar trend, but the $H_2$-1s peak is much narrower because of weak H-Au hybridization. The charge rearrangement in Fig. 4(b) indicates that $H_2$ transfers some electrons to the Au tip and no significant interaction occurs with the Au(110) substrate even for d=4.2 Å. Consequently, smooth variation of $E_b$ with d is obtained for $H_2$ in the tt-cavity.

In conclusion, systematic STM-IETS and DFT studies of a single $H_2$ molecule trapped in the tunable nanocavities formed in the STM junction between the tip and the Au (110) 2×1 reconstructed surface revealed substantial differences of the H-H bond and H-Au interactions in the tip-atop and tip-trough cavities. The vibrational and rotational energies are found to be highly sensitive to the tip-substrate distance for hydrogen trapped in the tip-atop cavity. This sensitivity also explains the enhanced spatial resolution of the rows of Au atoms at the surface in STM topographic images. By trapping and manipulating a $H_2$ molecule in the tip-substrate nano-cavity of the STM, the bond length and the rotational and vibrational energies of $H_2$ can be tuned and observed by changing the cavity dimension at a sub-Angstrom length scale. The comparison with DFT data for the geometric and spectroscopic features allows establishment of microscopic understanding of the chemical process in the tunable nanocavities. Such precise control and the availability of clear insights provide the opportunity to explore chemical processes with ultrahigh spatial resolution.

AUTHOR INFORMATION:
Corresponding Authors:
*Email: wilsonho@uci.edu; wur@uci.edu

Notes:

The authors declare no competing financial interest.

ACKNOWLEDGMENTS:

Work at UCI was supported by the National Science Foundation Center for Chemical Innovation on Chemistry at the Space-Time Limit (CaSTL) under Grant No. CHE-0802913 (H. W., S. L., H. H., A. Y.) and by the Chemical Sciences, Geosciences, and Biosciences, office of Basic Energy Science, U.S. Department of Energy, under Grant No. DE-FG02-06ER15826 (F. T., Z. H.). Work in Fudan was supported by Chinese National Science Foundation under grant 11474056 and the 1000 talent program of China (H. W., R. W.). Computer simulations were performed with the NSF Supercomputer Facility (through XSEDE).